%% file: aguchapter.tex
\documentclass[oneside,english,11pt]{article}
\usepackage[T1]{fontenc}
\usepackage[latin9]{inputenc}

\usepackage{import}
\usepackage{babel}
\usepackage{amsmath}
\usepackage{amsfonts}
\usepackage{amssymb}
\usepackage{graphicx}
\usepackage{xspace}
\usepackage{upgreek}
\usepackage[left=2cm,right=2cm,top=2cm,bottom=2cm]{geometry}
\usepackage{siunitx}

\import{includes/}{newcommands}
\renewcommand{\vec}{\boldsymbol}

\newcommand{\newfig}[3]{
 \begin{figure}
     \centering
     \includegraphics{fig_#1.pdf}
     \caption{#2}
     \label{#3}
 \end{figure}
}

\newcommand{\fig}{Figure }
\newcommand{\figs}{Figures }
\newcommand{\eq}{equation }
\newcommand{\eqs}{equations }

\newcommand{\citep}[1]{\cite{#1}}

\begin{document}

\title{Emergence of magnetic structure in supersonic isothermal magnetohydrodynamic turbulence}

\author{Jean-Mathieu Teissier$^{1,}$\footnote{Corresponding Author: jm.teissier@astro.physik.tu-berlin.de \\ $^1$ Zentrum f\"ur Astronomie und Astrophysik, Technische Universit\"at Berlin, D-10623, Berlin, Hardenbergstr. 36a, ER3-2, Germany \\ $^2$Max-Planck/Princeton Center for Plasma Physics}~~and Wolf-Christian M\"uller$^{1,2}$}

\date{12.11.2023}

\maketitle

\begin{abstract}
The inverse transfer of magnetic helicity is a fundamental process which may explain large scale magnetic structure formation and sustainement. Until very recently, direct numerical simulations (DNS) of the inverse transfer in magnetohydrodynamics (MHD) turbulence have been done in incompressible MHD or at low Mach numbers only. We review first results obtained through DNS of the isothermal MHD equations at Mach numbers ranging from subsonic to about 10. The spectral exponent of the magnetic helicity spectrum becomes flatter with increasing compressibility. When considering the Alfv\'en velocity in place of the magnetic field however, results found in incompressible MHD, including a dynamic balance between shear and twist, can be extended to supersonic MHD. In the global picture of an inverse transfer of magnetic helicity, three phenomena are at work: a local direct transfer mediated by the large scale velocity field, a local inverse transfer mediated by the intermediate scale velocity field and a nonlocal inverse transfer mediated by the small scale velocity field. The compressive part of the velocity field is geometrically favored in the local direct transfer and contributes to the nonlocal inverse transfer, but plays no role in the local inverse transfer.

Keywords: Magnetic helicity, Compressible, Supersonic, Turbulence, Magnetohydrodynamics, Direct numerical simulation
\end{abstract}

\newcommand{\MHIT}{ITMH\xspace}
\newcommand{\tdash}{\textemdash\xspace}
\newcommand{\gexpo}{g}

\newcommand{\mycitep}[1]{\citep{#1}}

\section{Introduction}

Flows, especially at geophysical or astrophysical scales, are mostly turbulent, so that intricate nonlinear interactions lead to chaotic behavior. In three-dimensional hydrodynamic turbulence, the large scale kinetic energy cascades to smaller and smaller scales through the break-up of large eddies into smaller ones, which successively experience a similar process, until viscous effects dissipate this energy into heat. In this sense, turbulence can be seen as a mechanism ``destroying'' continuously large-scale coherence, that is, the ``structure'' of the flow.

Despite this, large-scale magnetic structures embedded in turbulent plasmas are observed throughout the universe. They exist for example in and around planets with liquid core, stars, galaxies and clusters of galaxies \mycitep{BRS05}. It is thus important to understand the physical mechanisms responsible for the generation and sustainment of large-scale magnetic structures.

Two key processes in this respect are the dynamo and the inverse transfer of magnetic helicity (\MHIT in the following). ``Inverse transfer'' refers here to a transfer from small to large scales, in opposition to the ``direct cascade of kinetic energy'' in 3D hydrodynamics described above. In Fourier space, this corresponds to a transfer from large wavenumbers to small ones.

In a dynamo, a weak seed magnetic field is amplified through the conversion of kinetic to magnetic energy until nonlinear saturation. Dynamos are thought to be e.g. at the heart of planetary and stellar magnetospheres \citep{BRS05}.

While crucial in dynamo theory \citep{BLA15,VIC01} helical magnetic fields also have merits on their own. They generate large-scale structures without requiring conversion of kinetic to magnetic energy, but rather trigger a transfer of energy from the magnetic to the kinetic reservoir. The magnetic helicity $\mhel=\volavg{\FmagA \cdot \Fmag}$, with $\volavg{\cdot}$ denoting a volume average over the whole system, $\FmagA$ the magnetic vector potential and $\Fmag=\nabla \times \FmagA$ the magnetic field is a purely magnetic quantity which is subject to an inverse transfer. The theoretical possibility of \MHIT has first been established through absolute equilibrium statistical models in \cite{FPL75} and rapidly confirmed through diverse numerical experiments.

Magnetic helicity is a quadratic ideal invariant of the induction equation \citep{ELS56,WOL58a}:

\begin{equation}
	\partial_t \Fmag=\nabla \times (\Fvel \times \Fmag).
\end{equation}

The proof of invariance does not require the velocity field $\Fvel$ to be incompressible, not even to be an actual solution of, e.g., the magnetohydrodynamics equations \citep{ALU17}. Magnetic helicity is gauge invariant in the setting considered in this chapter (a closed periodic domain without mean magnetic field \citep{BER97}). In the general case, unless the volume is closed and the magnetic field component normal to the boundary vanishes, magnetic helicity is not gauge invariant. In such situations, one resorts to the so-called ``relative helicity'' \citep{BEF84}.

Magnetic helicity quantifies topological aspects of the magnetic field lines, such as their degree of linkage, twist, writhe and knottedness \citep{MOF69}. It is very well conserved even when low resistivity is present, e.g. in magnetic reconnection, significantly better than the total energy \citep{BER99}. In many geophysical and astrophysical plasmas, where low resistivity is typical, the quasi-conservation of magnetic helicity imposes strong constraints on the magnetic field dynamics. Apart from constraints on the geodynamo, magnetic helicity conservation is thought of playing a crucial role in e.g. solar flares and coronal mass ejections \citep{KUR96,LOW94}.

Most direct numerical simulations (DNS) of the \MHIT up to the present day have been performed in the incompressible case \citep{AMP06,LSM17,MFP81,MMB12,POP78,SFM15}. This choice is mostly made for the sake of conceptual and numerical simplicity:
\begin{itemize}
	\item on the conceptual side: incompressibility assumes a constant and uniform density, which is equivalent to the strict solenoidality of the velocity field. In that case, a degree of freedom for the velocity field is removed (in Fourier space, the velocity modes are in the plane orthogonal to their wavevector $\vk$, i.e. the available manifold for each mode is not 3D but 2D). The thermal pressure is thus reduced to a passive quantity, since one can solve for the vorticity instead of the velocity. Also, quadratic invariants are conserved in each nonlinear Fourier triad. However, in compressible turbulence, the situation is significantly more difficult. Some well-known theoretical results in incompressible turbulence have been extended to the compressible case only recently. For example, the existence of an inertial range for hydrodynamic compressible turbulence has been rigorously proven only in the last decade \citep{ALU13}. Similarly, considerations about e.g. an extension of or a variance from the incompressible Kolmogorov $-\frac{5}{3}$ spectral power law for velocity fluctuations in compressible flows have been derived theoretically and verified through numerical experiments around that same year \citep{FED13,GAB11}, as well as considerations about a compressible analogue to Kolmogorov's four-fifths law \citep{KWN13}.
	\item on the numerical side: incompressible flows with periodic boundary conditions can very efficiently be dealt with by pseudospectral methods \citep{CHQ88}. Supersonic compressible flows on the other hand lead to shocks and discontinuities, for which a Fourier representation is not suitable (e.g. because of the Gibbs phenomenon). Reconstructing abruptly varying quantities naively through polynomial interpolation on uniform grids results in strong oscillations. Resolving such cases properly without sacrificing too much accuracy in smooth regions required the advent of higher-order shock-capturing methods able to adjust themselves depending on whether the zone considered presents a shock or not. The first Essentially Non-Oscillatory (ENO) schemes, of discretisation order higher than two and able to adapt to discontinuities, were designed end of the 80s and then extended to the more stable Weighted ENO (WENO) schemes in the 90s \citep{HEO87,JIS96,LOC94,SHU09}. High-order time integration techniques avoiding the addition of oscillations as part of the time-stepping procedure were developped about 15 years ago \citep{KET08}. Techniques allowing a numerically more efficient dimension-by-dimension approach have been designed a couple of years later \citep{BUH14,COC11}. For very strong shocks, maintaining the positivity of the mass density and the thermal pressure is not a trivial matter, and is still an active area of research \citep{BAL12,BGS16,WSH19}. Thus, viable numerical schemes allowing high accuracy supersonic DNS simulations with acceptable numerical costs are relatively recent. Consequently, the few DNS studies using an adiabatic or an isothermal equation of state have been performed in the subsonic or transonic case \citep{BAP99,BRA01,CHB01}.
\end{itemize}

Incompressibility is however of limited applicability in many astrophysical situations, where high Mach numbers (the ratio of the flow speed to the speed of sound) are common. For example, the interstellar medium exhibits root mean square (RMS) turbulent Mach numbers typically ranging from about 0.1 to  10 (\cite{ELS04}, section 4.2).

In this chapter, we review some results of the \MHIT in incompressible magnetohydrodynamics (MHD) and first results of the \MHIT in compressible supersonic MHD, obtained through fourth-order accurate finite-volume DNS of the isothermal ideal equations (i.e. with neither physical viscosity nor resitivity) \citep{TEM21a,TEM21b}. In the presence of a mechanical and an electromotive forcing, they read:

\beqa
	\label{eq:MHDden} \pat \rho&=&-\nabla \cdot (\rho \Fvel),\\
	\pat (\rho \Fvel)&=&- \nabla \cdot \left( \rho \Fvel \Fvel^T + (\rho \cs^2 + \frac{1}{2}|\Fmag|^2)\mathI - \Fmag \Fmag^T \right)+\rho\ForceV,\\
	\pat \Fmag&=&\nabla \times (\Fvel \times \Fmag)+\ForceM,
\eeqa

with the solenoidality constraint:

\beq
	\label{eq:MHDdivb} \nabla \cdot \Fmag=0.
\eeq

The symbols have their usual meaning: $\rho$ is the mass density, $\Fvel$ the velocity, $\cs$ the constant sound-speed in the isothermal case so that the pressure is $p=\rho \cs^2$ and $\Fmag$ is the magnetic field. The kinetic forcing $\ForceV$ takes place through an acceleration field, $\ForceM$ is the electromotive driving and $\mathI$ is the $3\times 3$ identity matrix. The equations above are written in conservative form, as appropriate when using a finite-volume solver. The MHD approximation describes well the dynamics of a conducting fluid, such as planetary fluid metallic cores or astrophysical plasmas at large enough time and space scales \citep{KIV95}.
 
The isothermal equation of state is a reasonable approximation for example in the early homologous phase of a gravitational collapse of a molecular cloud, where the heat generated through the collapse is removed by radiative cooling. During a later phase, the plasma becomes optically thick and the temperature rises in a nearly-adiabatic way (\cite{CAO07}, section 12.2). 

Thus, the isothermal approximation can be considered as the next step of complexity when departing from strict incompressibility. It introduces one new variable: a non-constant density, associated with the internal energy density $u=\rho \cs^2 \ln(\rho/\rhoz)$, with $\rhoz$ the mean density of the system \citep{KWN13}. The variable density allows for finite $\nabla \cdot \Fvel$.

The main aims of the investigations of supersonic isothermal MHD flows \citep{TEM21a,TEM21b} reviewed in this chapter is to explore similarities and differences between scaling laws in incompressible and supersonic MHD turbulence. The role of the compressive velocity field is also in focus. For this, DNS without mean magnetic field under periodic boundary conditions are performed, so that the magnetic helicity is gauge invariant. For the sake of simplicity, only one sign of magnetic helicity is injected. Although it is not physically motivated, since both signs are expected in nature, the results in this already complex setting should be helpful to interpret more realistic configurations in the future.

This chapter is structured as follows. The spectral analysis tools used, which allow to disentangle helical components and obtain information about locality in Fourier space, are presented in section \ref{sec:anatools}. The results of the isothermal supersonic DNS are presented in section \ref{sec:results}, after a review of some previously established results in incompressible MHD turbulence (section \ref{sec:prevresearch}) and a presentation of the numerical method used (section \ref{sec:solver}). Concluding remarks are finally given in section \ref{sec:conclusion}.

\section{Fourier diagnostics, shell-to-shell and helical transfers}
\label{sec:anatools}
\subsection{Shells, spectra and co-spectra}

In the chosen setup with periodic boundary conditions in a cubic box of size $\Lbox$, each field $\neutrfield$ can be represented as a discrete sum of its Fourier harmonics, with Fourier coefficients $\neutrfieldF_\vk$ at wavevector $\vk$. The power spectrum $\powerspec(\neutrfield)$ is thus computed through a summation over Fourier shells $S_\kspec=\{\vk | \kspec\leq |\vk|/\kone <\kspec+1\}$, with $\kspec\in\mathbb{N}$ and $\kone=\frac{2\upi}{L}$ the smallest wavenumber in the system:

\beq
	\label{eq:fshellFK}
	\powerspec(\neutrfield)_\kspec=\frac12 \sum_{\vk \in S_\kspec} |\neutrfieldF_\vk|^2=\frac12 \volavg{|\neutrfield_\kspec|^2},
\eeq

where the second equality comes from Parseval's theorem, with $\neutrfield_\kspec$ the inverse Fourier transform of the shell-filtered $\neutrfield$ at shell $\kspec$ (i.e. keeping only $\neutrfieldF_\vk$ where $\vk\in S_\kspec$ and setting the other Fourier coefficients to 0).

Similarly, the co-spectrum of two fields $\neutrfield, \neutrfieldB$ is defined by:

\beq
	\productspec(\neutrfield,\neutrfieldB)_\kspec= \sum_{\vk \in S_\kspec} \neutrfieldF_\vk \cdot \neutrfieldBF_\vk^*= \volavg{\neutrfield_\kspec \cdot \neutrfieldB_\kspec},
\eeq

where the star in the superscript denotes the complex conjugate. The magnetic helicity spectrum is thus $\mhelF=\productspec(\FmagA,\Fmag)$.

\subsection{Helical triadic interactions}
The decomposition of a field in helical components is based on the diagonalisation of the curl operator in Fourier space $i\vk\times\cdot$. This operator possesses the eigenvalues $(0,+k,-k)$, with $k=|\vk|$, associated respectively with the unitary eigenvectors $(\vhhk,\vhhp,\vhhm)$ where $\vhhk=\vk/|\vk|$ and (\cite{WAL92}, see also \cite{BRS05}):

\beq
	\label{eq:vhhpm}
    \vhhpm = \frac{1}{\sqrt{2}}\frac{\vk \times (\vk \times \vhe) \mp i k (\vk \times \vhe)}{k^2\sqrt{1-(\vk \cdot \vhe/k)^2}},
\eeq

with $\vhe$ an arbitrary unitary vector non-parallel to $\vk$.

This decomposition can be seen as an extension of the Helmholtz decomposition in compressive (parallel to $\vk$) and solenoidal (orthogonal to $\vk$) components. Indeed, the plane orthogonal to $\vk$ is spanned by the eigenvectors $\vhhpm$, corresponding to circularly polarised waves with opposite polarity. In configuration space, flow lines corresponding to these eigenvectors form a left- or right-handed helix.

A rigorous mathematical framework for the helical decomposition has been introduced by \cite{COM88}  (where it is called ``Beltrami decomposition''). In three-dimensional hydrodynamic turbulence, this decomposition allows to find helical interactions responsible for a subdominant inverse transfer in the global picture of a direct cascade of energy \citep{BMT12,WAL92}, without the need of rotation or confinement (quasi-2D flow). This may be relevant for turbulence models in large-eddy simulations \citep{ALE17}. On a mathematical level, a helical decimation of the Navier Stokes equations (i.e. their projection on the positive helical eigenvectors) gives some insight regarding e.g. the existence and uniqueness of weak solutions \citep{BIT13}. Extensions to the MHD case may provide information to disentangle the intertwined dynamics of magnetic and kinetic helicities \citep{ALB18,LPC09,LBM16,LID17,LSM17}.

A magnetic field mode can thus be represented as $\FmagF_{\vk}=\sum_{\hsk\in\{+,-\}}\FmagFL^{\hsk}_{\vk}\vhhs{\hsk}{\vk}$, where $\hsk=0$ is not allowed since the magnetic field is solenoidal. In contrast, a velocity field mode can have a compressive component so that $\FvelF_{\vp}=\sum_{\hsp\in\{+,0,-\}}\FvelFL^{\hsp}_{\vp}\vhhs{\hsp}{\vp}$.

Using this decomposition, the time variation of helical component $\FmagFL^{\hsk}_{\vk}$ of the magnetic field is given in incompressible MHD by \citep{LPC09}:

\beq
  \label{eq:pathelb}
  \pat \FmagFL^{\hsk}_{\vk}=\hsk k \sum_{\vk+\vp+\vq=0} \sum_{\hsp,\hsq} \FvelFL^{\hsp*}_{\vp} \FmagFL^{\hsq*}_{\vq} \triadgeomkpq.
\eeq

This relation remains valid in compressible MHD, the only difference being that $\hsp\in\{+,0,-\}$ and not only $\hsp=\pm$ any longer. The geometric factor $\triadgeomkpq$ depends on the triad shape and the helical components considered. Its modulus quantifies the importance of the respective triad interaction and is given by \citep{WAL92}:

\beq
  \label{eq:Gincomp}
  G^{\hsk,\hsp \in \{+,-\},\hsq}_{k,p,q}=\frac{|\hsk k+\hsp p+\hsq q|\sqrt{2k^2p^2+2p^2q^2+2q^2k^2-k^4-p^4-q^4}}{2kpq},
\eeq

for the incompressible modes and by:

\beq
  \label{eq:Gcomp}
  G^{\hsk,\hsp=0,\hsq=\pm \hsk}_{k,p,q}=\frac{|(q\mp k)(p^2-k^2-q^2 \mp 2qk)|}{2kpq}
\eeq	

when the velocity field component considered is the compressive one \citep{TEM21b}.

\subsection{Shell-to-shell transfers}

Contrary to the energy, which can be transferred from the kinetic energy reservoir to the magnetic one and vice-versa, magnetic helicity is a purely magnetic quantity. The velocity field present in the triadic interactions (see \eq \eqref{eq:pathelb}) only plays the role of a mediator. Summing over all the wavenumbers present in a certain shell $\kspec$, one can derive the transfer of magnetic helicity from a shell $Q$ to a shell $K$, mediated by the velocity field shell $P$ \citep{AMP06,PSV19}:

\beq
	\label{eq:tsfmhlqpk}
	\TsfMhlQPK=2\volavg{\FmagK \cdot (\FvelP \times \FmagQ)},
\eeq	

with the subscripts on $\Fmag$ and $\Fvel$ denoting at which shell the fields have been filtered (see \eq \eqref{eq:fshellFK}). We remind that the volume averaging $\volavg{\cdot}$ takes place in configuration space. The derivations of \citep{AMP06} done in the incompressible case remain valid in compressible MHD because of the magnetic field solenoidality. The interpretation as a transfer rate of magnetic helicity between shells is consistent with the antisymmetric property $\TsfMhl(Q,P,K)=-\TsfMhl(K,P,Q)$. Then, in the absence of an electromotive driving and of dissipation, the magnetic helicity present in shell $\kspec$, $\mhelFK=\volavg{\FmagAK \cdot \FmagK}$, is governed by:

\beq
  \label{eq:tsfmhlqk}
  \pat \mhelFK = \sum_Q \sum_P \TsfMhlQPK.
\eeq

To disentangle the role of the different helical components, the $\Fvel$ and $\Fmag$ filtered fields can first be projected on the helical eigenvectors $(\vhhk,\vhhp,\vhhm)$, giving \citep{TEM21b}:

\beq
  \label{eq:tsfmhlhel}
  \TsfMhlXXX{\hsK\hsP\hsQ}(Q,P,K)=2\volavg{\FmagXX{\hsK}{K} \cdot (\FvelXX{\hsP}{P} \times \FmagXX{\hsQ}{Q})},\quad\quad\pat \mhelFK = \sum_Q \sum_P \sum_{\hsK,\hsP,\hsQ} \TsfMhlXXX{\hsK\hsP\hsQ}(Q,P,K).
\eeq

In the above expression $\FmagXX{\hsK}{K}$ represents the positively (for $\hsK=+$) or negatively ($\hsK=-$) helical magnetic field (i.e. projected on $\vhhs{\hsK}{}$), filtered at shell $K$. The last summation on the right takes place over 12 terms: $\hsK,\hsQ \in \{+,-\}, \hsP\in \{+,0,-\}$. The terms where $\hsK=\hsQ$, labelled here as ``homochiral'' are antisymmetric and can be interpreted as magnetic helicity shell-to-shell transfers depending on the helical components. However, the terms where $\hsK=-\hsQ$ have to be considered in pairs to exhibit the antisymmetric property: $\TsfMhlXXX{\hsK\hsP\hsQ}+\TsfMhlXXX{\hsQ\hsP\hsK}$. In the considered investigations, these ``heterochiral'' terms have a relatively low magnitude and are therefore not treated in much detail. 

\section{Results from previous research in incompressible MHD}
\label{sec:prevresearch} 

The first prediction regarding the magnetic helicity spectral exponent has been obtained through a dimensional analysis ``\`a la Kolmogorov'' \citep{PFL76}. Assuming local transfers and a constant flux of magnetic helicity in an inertial range, a dimensional analysis gives $[\FmagA][\Fmag]/T=[\Fmag]^2[\Fvel]\sim const.$ In the incompressible case, $[\Fvel]\sim[\Fmag]$ so that $b_l^3 \sim (\mhelF_l/l)^{3/2}\sim const.$, where the subscript $l$ means typical values for eddies of size $l$. This leads to $\mhelF_l \sim l$ so that for the magnetic helicity spectrum $\mhelFK \sim \kspec^{-2}$. The spectral exponent of $-2$ has been observed in numerical simulations of the EDQNM (Eddy-Damped Quasi Normal Markovian) model extended to MHD \citep{PFL76}.

However, later DNS of incompressible MHD and Lagrangian-averaged MHD have shown a significantly steeper spectral exponent of the $\mhelF$ spectrum, in the range $[-3.7,-3.3]$ \citep{GMP11,MIP09,MMB12}. This echoes with the word of caution announced in \cite{PFL76}, warning that the $-2$ scaling may be altered by nonlocal effects. Indeed, \cite{AMP06} have shown through Fourier shell-to-shell analysis that the \MHIT presents both local and nonlocal features.

Based on the EDQNM closure model, a dynamical equilibrium between shearing and twisting effects leads to the balance (\cite{MUM10}, see also \cite{MIP09,GMP11,MMB12}):
\beq
	\label{eq:Alfbalance}
	\Big(\frac{\sekinF}{\emagF}\Big)^{\gexpo} \propto \frac{\khelF}{\jhelF},
\eeq

with $\sekinF=\powerspec(\Fvel)$ and $\emagF=\powerspec(\Fmag)$ the (specific) kinetic and magnetic energy spectra, $\khelF=\productspec(\Fvel,\vom)$ and $\jhelF=\productspec(\Fmag,\nabla \times \Fmag)\approx(\frac{2\upi}{L}K)^2\mhelF$ the kinetic and current helicity spectra and $\gexpo$ a constant exponent. This balance has been verified empirically with an exponent $\gexpo=1$ in decaying turbulence experiments, with large-scale helical magnetic fields \citep{GMP11,MIP09}. For the incompressible investigation of the \MHIT \citep{MUM10}, the data is also compatible with an exponent $\gexpo=1$, even though later analysis have shown that an exponent $\gexpo=2$ fits the data better \citep{MMB12}, presumably because of different domains where the individual spectra $\sekinF,\emagF,\khelF$ and $\jhelF$ present scaling laws. In \cite{GMP11} this equilibrium has been interpreted as a ``partial Alfv\'enization of the flow'', so that we refer to relation \eqref{eq:Alfbalance} as the ``Alfv\'enic balance''.
 
Regarding the strength and locality of the \MHIT, a shell-to-shell analysis in incompressible MHD \citep{AMP06} shows that the transfer is more local at early times and nonlocal at later times, when magnetic helicity is present at the largest scales. This study shows as well the presence of a local direct transfer of magnetic helicity. Furthermore, using a helical decomposition in the framework of linear stability analysis, \cite{LSM17} have shown that transfers between two like-signed helical magnetic modes (e.g. from $b_{\vq}^+$ to $b_{\vk}^+$) are stronger and more nonlocal when mediated by a like-signed helical velocity mode (i.e. $v_{\vp}^+$), as compared to an oppositely-signed one ($v_{\vp}^-$).

These results in incompressible MHD are put in relation with those in the supersonic isothermal case \citep{TEM21a,TEM21b} in section \ref{sec:results}.

\section{Numerical method}
\label{sec:solver}

The results in supersonic MHD reviewed here \citep{TEM21a,TEM21b} have been obtained through DNS of the 
ideal isothermal MHD equations \eqref{eq:MHDden}-\eqref{eq:MHDdivb}. The finite-volume shock capturing numerical solver used (\cite{VTH19}, see also \cite{TEI20}) is fourth-order accurate in space and time, ensures magnetic field solenoidality through the constrained transport method and resorts to lower order reconstruction in the vicinity of strong discontinuities (fallback approach) to keep the mass density positive at high Mach numbers.

\newfig{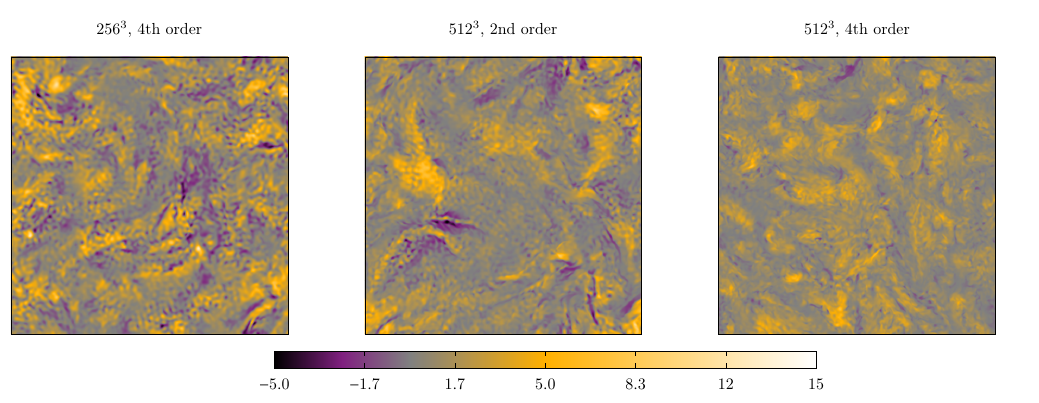}{Slices of the magnetic helicity density, normalized by its mean in the system, for three runs, at an instant when the magnetic helicity integral scale $\IscHm\approx\Lbox/6$. These runs correspond to a Mach number close to unity in the statistically stationary state and are done at different resolutions and using numerical schemes of different discretization order.}{fig:lrlo}

The use of a fourth-order numerical code allows results of good accuracy at the moderate $512^3$ resolution (see a comparison with a second-order solver in \fig \ref{fig:lrlo}). The main results are confirmed at resolution $1024^3$.

Initially, the plasma is at rest ($\rho=\rho_0=1$ is constant and $\Fvel=0$) in a triply periodic cubic box $[0,\Lbox=1]^3$ and no magnetic field is present. The isothermal sound speed is taken as $\cs=0.1$. An Ornstein-Uhlenbeck mechanical forcing $\fOUF$ injects kinetic energy at the largest scales with a specified degree of compressivity. The forcing is defined in Fourier space, initialized with $\fOUF=0$ and evolved in time according to:

        \beq
                \label{eq:dtfOUF}
                d\fOUFk(t)=-\fOUFk(t)\frac{dt}{\fat}+\famp \left( \frac{2\sigma(\vk)^2}{\fat} \right)^{1/2} \projkT \cdot d\vW(t),
        \eeq

with $\sigma(\vk)=1$ for the largest scales $\vk \in S_1 \cup S_2$ and $\sigma(\vk)=0$ otherwise. The autocorrelation time is set to an estimate of the turbulent turnover time: $\fat=\Lbox/(2\cs\rmsM^*)$, with $\rmsM^*$ the expected stationary state root mean square (RMS) Mach number. A three-dimensional continuous random walk is performed via the Wiener process $d\vW(t)=dt\vN(0,dt)$, with $\vN(0,dt)$ a 3D Gaussian distribution with zero mean and standard deviation $dt$. The compressiveness of the forcing is determined by the projector $\projkij(\vk)=\specw \delta_{ij}+(1-2\specw) \frac{k_ik_j}{|k|^2}$. Only the extreme cases $\specw=1$ (purely solenoidal forcing) and $\specw=0$ (purely compressive forcing), cf. \cite{FED13}, are considered. The field $\fOUF$ is back-transformed in configuration space and normalized at each timestep so that a constant energy injection rate $\Einj$ is achieved \citep{TEM21a}. As a consequence, the amplitude $\famp$ can be chosen arbitrarily. The finite-volume formalism guarantees the conservation of momentum, but not of mean velocity. Hence, the weak fluctuating mean velocity field is removed at each iteration, as the presence of magnetic fields breaks Galilean invariance.

The kinetic energy cascades to smaller and smaller scales. At the smallest scales, numerical dissipation is dominant so that a hydrodynamic statistically stationary state is reached with a desired degree of compressivity and RMS Mach number $\rmsM$, determined by both $\Einj$ and $\specw$. In order to cover typical Mach numbers found in the interstellar medium, the following runs have been performed:

\begin{itemize}
\item with a purely solenoidal driving: M01s, M1s, M5s, M7s, M11s where the number represents the average RMS turbulent Mach number in the statistically stationary state, respectively $\rmsM \approx 0.12, 1.1, 5.1, 7.0, 11.$,
\item with a purely compressive driving: M1c, M3c, M5c, M8c, with $\rmsM \approx 0.80, 2.8, 5.1, 7.9$ respectively.
\end{itemize}

\newfig{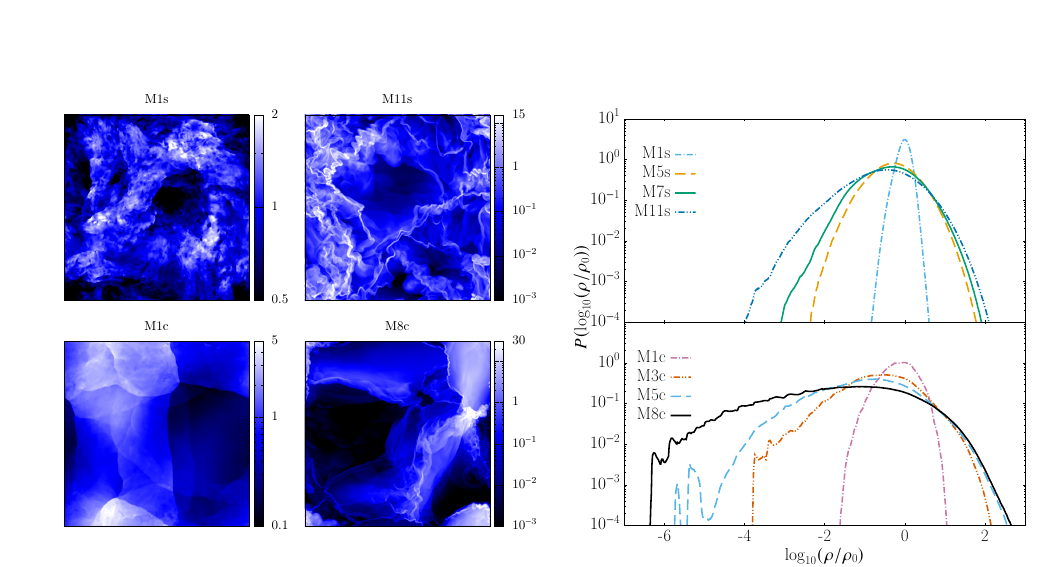}{Mass density properties during the hydrodynamic statistically stationary state. Left: mass density slices for the M1s, M11s, M1c and M8c runs. Right: Probability distribution function of $\log(\rho/\rhoz)$ averaged through roughly one turbulent turnover time.}{fig:denstats}

The mass density $\rho$ exhibits an increasing amount of filamentary structures with increasing $\rmsM$. The compressively-driven runs, with a high $\nabla \cdot \Fvel$ component, have significantly stronger mass density variations as compared to the solenoidaly-driven ones. This is made apparent through large regions of very low density and comparison of the probability distribution functions of $\log(\rho/\rhoz)$ (\fig \ref{fig:denstats}). The compressively-driven runs also present pronounced shock fronts, already at Mach numbers of the order of unity \citep{TEM21a}.

At a chosen instant in the statistically stationary state, an electromotive forcing at small scales is switched on, while the large-scale mechanical forcing is maintained. The electromotive forcing is a delta-correlated (white noise in time), maximally positive helical driving, i.e. with all the components along the $\vhhp$ eigenvector (see \eq \eqref{eq:vhhpm}). Generated in Fourier space for the shells $\bigcup_{48\leq \kspec \leq 52} S_\kspec$, it is back-transformed and normalized in configuration space so that a constant magnetic energy injection rate $\EMinj$ is achieved. For all the runs, $\EMinj=\Einj$, apart for M01s and M1s where it is $4\Einj$ and $2\Einj$ respectively in order to reach faster the convergence to a power-law behavior of the magnetic helicity spectrum \citep{TEM21a}. As verified numerically, the cross-helicity remains low for all the runs so that the dynamics are dominated by the direct cascade of energy and the \MHIT.

The simultaneous forcing at two scales might seem odd at first sight, but it is necessary in order to observe the effects of the compressive velocity field. Indeed, its effects enter the magnetic field dynamics through a term $\sim \nabla \cdot \Fvel$. Due to the direct cascade of kinetic energy, only a large-scale driving can generate a dynamically important compressive velocity field in the range of scales  characteristic of the inverse transfer of magnetic helicity. Secondary velocity fluctuations generated by Alfv\'enic oscillations of the growing magnetic structures are not sufficient.

The injected magnetic helicity is subsequently transferred to ever larger scales. Since there is no large scale energy sink, the asymptotic state of the system would be a condensate of magnetic helicity at the largest scales, which would dominate the dynamics. In particular, this finite-size effect would strongly affect the locality of transfers and the spectral exponents \citep{AMP06,LID16}. Hence, the results presented here consider an instant in time when an energetically dominant largest-scale magnetic structure has not yet emerged (more precisely, when the magnetic helicity integral scale $\IscHm=\Lbox(\int_{\kspec} \kspec^{-1} \mhelF d\kspec)/(\int_{\kspec} \mhelF d\kspec)$ is about $\Lbox/8$ or $\Lbox/6$). Although not ideal, this choice of stopping the runs earlier presents a pragmatic compromise between an asymptotic scaling range polluted by a large-scale sink of magnetic helicity and perfect statistical stationarity. This allows to observe dependencies of the spectral scaling of magnetic helicity and derived quantities on the Mach number and the forcing mechanism with limited numerical expense.

\section{Results}
\label{sec:results}

The next subsections \ref{sec:TEM21a} and \ref{sec:TEM21b} present the main results of  \citep{TEM21a,TEM21b}, respectively. Additional material and more detailed analysis can be found therein.

\subsection{Spectra and Alfv\'enic balance}
\label{sec:TEM21a}

\newfig{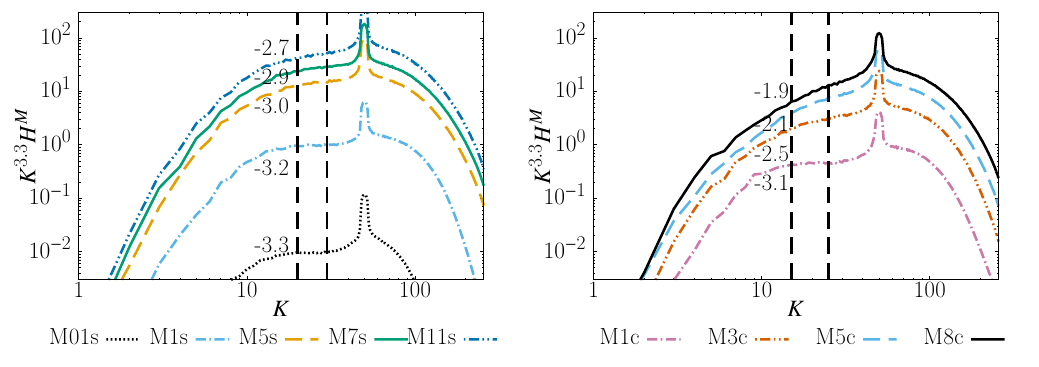}{Magnetic helicity spectra compensated by $K^{3.3}$ at an instant when $\IscHm \approx \Lbox/8$. The numbers indicate the spectral slope measured through a least-squares fit in the domain delimited by the two vertical dashed lines.}{fig:mhlspec}

The results presented in this subsection are taken at an instant when the magnetic helicity integral scale $\IscHm \approx \Lbox/8$ (cf section \ref{sec:solver}) so that an energetically dominant largest-scale magnetic structure has not emerged yet.

At low Mach number (for the M01s, M1s and M1c runs), the magnetic helicity spectra exhibit a scaling consistent with results found in incompressible MHD: $\mhelFK\sim K^m$ with $m\approx -3.3$ \citep{MMB12}. For increasing compressibility, the spectra become flatter ($|m|$ decreases), see \fig \ref{fig:mhlspec}. The scaling is significantly more dependent on the flow's compressivity (the $\nabla \cdot \Fvel$ component) than on the RMS Mach number. For example, deviations from the incompressible scaling are higher for the M3c run ($\mhelFK\sim K^{-2.5}$) than for the M11s one ($\mhelFK\sim K^{-2.7}$). Although the trend towards flatter spectra is very clear, the precise numerical values of the exponents presented here should be taken with appropriate caution, since only a particular snapshot is considered. For example, starting to inject magnetic helicity at a different instant in the hydrodynamic statistically stationary state of the M5c run, when the mass density PDF is peaked at lower values, results in a flatter scaling $\mhelFK\sim K^{-1.1}$. Furthermore, at a later instant in time, the departure of the most compressible solenoidally-driven runs from the incompressible case is smaller \citep{TEM21a}.

\newfig{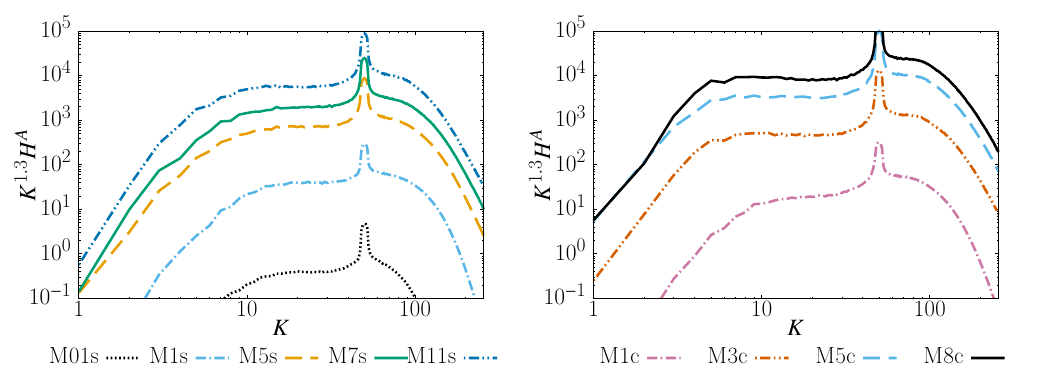}{Alfv\'enic helicity spectra compensated with $K^{1.3}$, at the same instant as in \fig \ref{fig:mhlspec}.}{fig:ahlspec}

In the course of the \MHIT, the magnetic field experiences topological changes which require magnetic reconnection processes. The changes are communicated through Alfv\'en waves which propagate at the Alfv\'en speed $|\FvelA|=|\Fmag/\sqrt{\rho}|$ along the affected magnetic field lines. In incompressible MHD, the magnetic field and the Alfv\'en velocity are the same up to a constant multiplicative factor. On the contrary, in the compressible case, low density regions are associated with shorter Alfv\'enic timescales $\propto 1/|\FvelA|$. This suggests to consider the Alfv\'en velocity in place of the magnetic field. The spectral ``Alfv\'enic helicity'' $\ahelF=\productspec(\FvelA,\nabla \times \FvelA)$, exhibits for all the runs considered a very similar scaling $\ahelFK \sim K^m$ with $m \approx -1.3$ (\fig \ref{fig:ahlspec}). This scaling is, for all the runs apart M1c, close to the one of the corresponding quantity in incompressible MHD: $\jhelF=\productspec(\Fmag,\nabla \times \Fmag)$, the current helicity, which scales as $\jhelFK\sim K^2 \mhelFK \sim K^{-1.3}$. The fact that the scaling exponent does not change significantly over a wide range of compressibility is an indicator for a systematic scale-dependent correlation between the magnetic field and the density. The M1c run may show different dynamics because the compressive ratio of the velocity field stays high in the inverse transfer region, contrary to the higher Mach number runs, where extended low density regions are strongly affected by the magnetic field.

\newfig{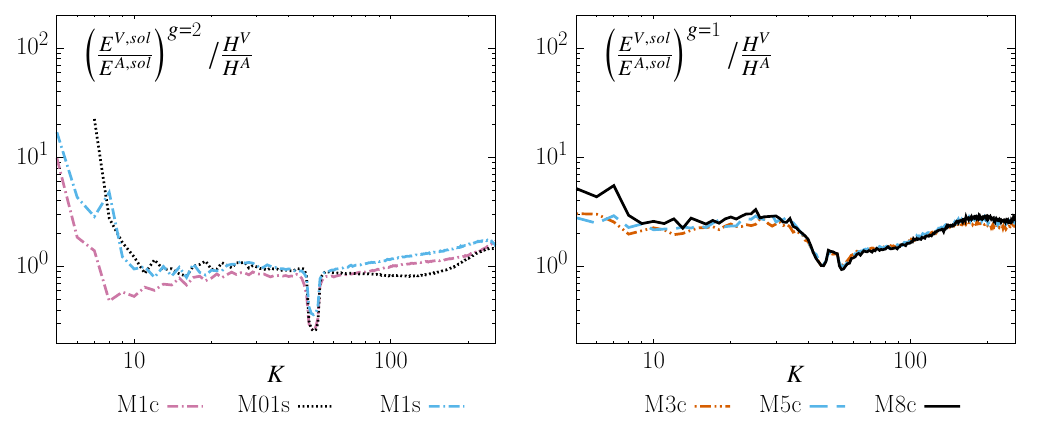}{Test of the Alfv\'enic balance with $\gexpo=2$ for the low Mach number runs (left) and $\gexpo=1$ for the most compressive runs (right). The same instant as in \fig \ref{fig:mhlspec} is considered.}{fig:alfbalance}

This density-magnetic field correlation suggests to extend the Alfv\'enic balance \eqref{eq:Alfbalance} by considering the Alfv\'en velocity in place of the magnetic field:

\beq
	\label{eq:AlfbalanceC}
	\Big(\frac{\sekinFsol}{\ealfFsol}\Big)^{\gexpo} \propto \frac{\khelF}{\ahelF},
\eeq

with $\ealfFsol=\powerspecsol(\FvelA)$ and $\sekinFsol=\powerspecsol(\Fvel)$ the power spectra obtained by considering only the solenoidal components of the fields (orthogonal to $\vk$ in Fourier space). This balance is well followed for the subsonic and transonic runs M01s, M1s and M1c with an exponent $\gexpo=2$ (\fig \ref{fig:alfbalance}), consistent with investigations in incompressible MHD \citep{MMB12}. For the most compressive supersonic runs M3c, M5c and M8c, it is well followed with an exponent $\gexpo=1$. The supersonic solenoidally-driven runs M5s, M7s and M11s display an intermediate behavior.

Relation \eqref{eq:AlfbalanceC} may be seen as the most straightforward extension of relation \eqref{eq:Alfbalance} for compressible flows, i.e. considering only the solenoidal part of the velocity field (the only one that is present in incompressible flows as well). Other possibilites have been tried, e.g. considering $\powerspec(\sqrt{\rho}\Fvel)$, $\frac{1}{2}\productspec(\Fvel,\rho \Fvel)$ among others, in place of $\sekinFsol$, but the best agreement giving the least spread of the curves has been obtained with relation \eqref{eq:AlfbalanceC}. This has also been confirmed at the higher $1024^3$ resolution.

The extension of the Alfv\'enic balance suggests that the shear, not the compressive waves, balances the twisting effects in supersonic isothermal MHD.

\subsection{Shell-to-shell transfers}
\label{sec:TEM21b}

\newfig{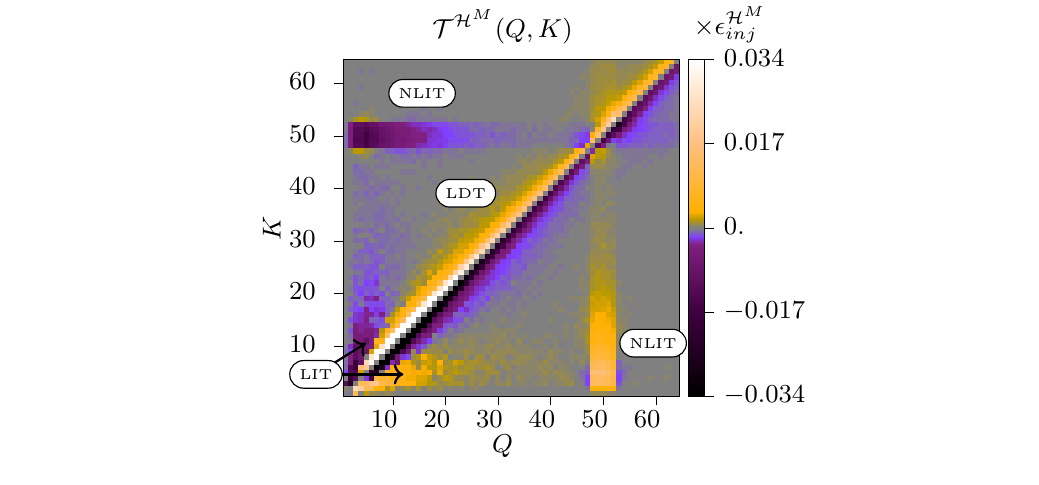}{Magnetic helicity transfer rates from shell $Q$ to shell $K$, $\TsfMhl(Q,K)=\sum_P \TsfMhl(Q,P,K)$ (cf \eq \eqref{eq:tsfmhlqpk}) for the M8c run at an instant when $\IscHm \approx \Lbox/6$. The closest (resp. furthest) to the main diagonal, the more local (resp. nonlocal) the transfer. Above the diagonal, a bright/orange color means a direct transfer (from large to small scales) and a dark/purple color an inverse transfer, and vice-versa under the diagonal. LDT stands for local direct transfer, LIT for local inverse transfer and NLIT for nonlocal inverse transfer.}{fig:tsfhmqk}

We review here shell-to-shell transfer results from the most compressive M8c run at an instant when $\IscHm \approx \Lbox/6$. The displayed transfer rates are in units of the magnetic helicity injection rate (or rather, an estimate thereof): $\mhelinj=(2\EMinj\mhelfracFor)/(2\upi\mhlkinj/L)$ with $\mhelfracFor=+1$ the helical fraction of the electromotive driving occuring around $\mhlkinj=50$. 

The shell-to-shell transfer analysis reveals three features in the global picture of the \MHIT (\fig \ref{fig:tsfhmqk}): $(i)$ a local direct transfer (LDT), $(ii)$ a local inverse transfer (LIT) and $(iii)$ a nonlocal inverse transfer (NLIT) from the electromotive forcing scale up to the largest ones. This is consistent with the findings in incompressible MHD \citep{AMP06}, which reports both a LDT and an inverse transfer: LIT at early times and NLIT at later times. In that investigation, the electromotive driving takes place around the shell $\kspec=8$ so that altough both LIT and NLIT are actually present at later times, they are more difficult to separate from one another.

Each of these three features can be associated with different spatial scales of the mediating velocity field. This is shown in \fig \ref{fig:tsfmediator}, which displays the transfer of magnetic helicity from shell $Q$ to shell $K$ mediated by a range of velocity field shells $P$. The LDT is associated with a direct cascade of magnetic energy and is mediated by the large-scale velocity field ($1\leq P \lessapprox 3$). The NLIT is mediated by the small scale velocity field ($30 \lessapprox P$) and can be interpreted as a merging of small-scale magnetic fluctuations with a much larger magnetic structure. The LIT corresponds to the merging of similar-sized magnetic structures (within a factor of about 2) and is mediated by the intermediate scales of the velocity field ($4 \lessapprox P \lessapprox 29$).

\newfig{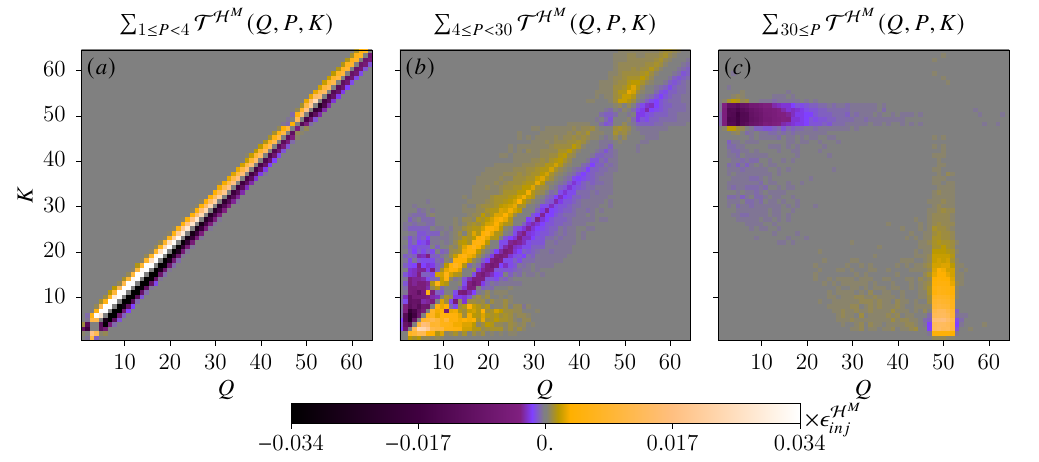}{Filtering of the transfer rates displayed in \fig \ref{fig:tsfhmqk} according to the velocity field shells $P$. Transfers mediated by: $(a)$ the large-scale velocity field, $(b)$ the intermediate scales and $(c)$ the small-scale velocity field.}{fig:tsfmediator}

The helical decomposition allows furthermore to associate the LDT, the LIT and the NLIT with different helical components of the velocity field. For the M8c run, the largest terms are the ones mediated by the helically like-signed velocity field, as well as its compressive part (\fig \ref{fig:tsfhmqkhel}): $\TsfMhlXXX{\PPP}$ and $\TsfMhlXXX{\PCP}$ (cf \eq \eqref{eq:tsfmhlhel}). The other helical transfer terms are relatively small and therefore not shown here (apart from $\TsfMhlXXX{\PNP}$, to which we refer in the next point).

\newfig{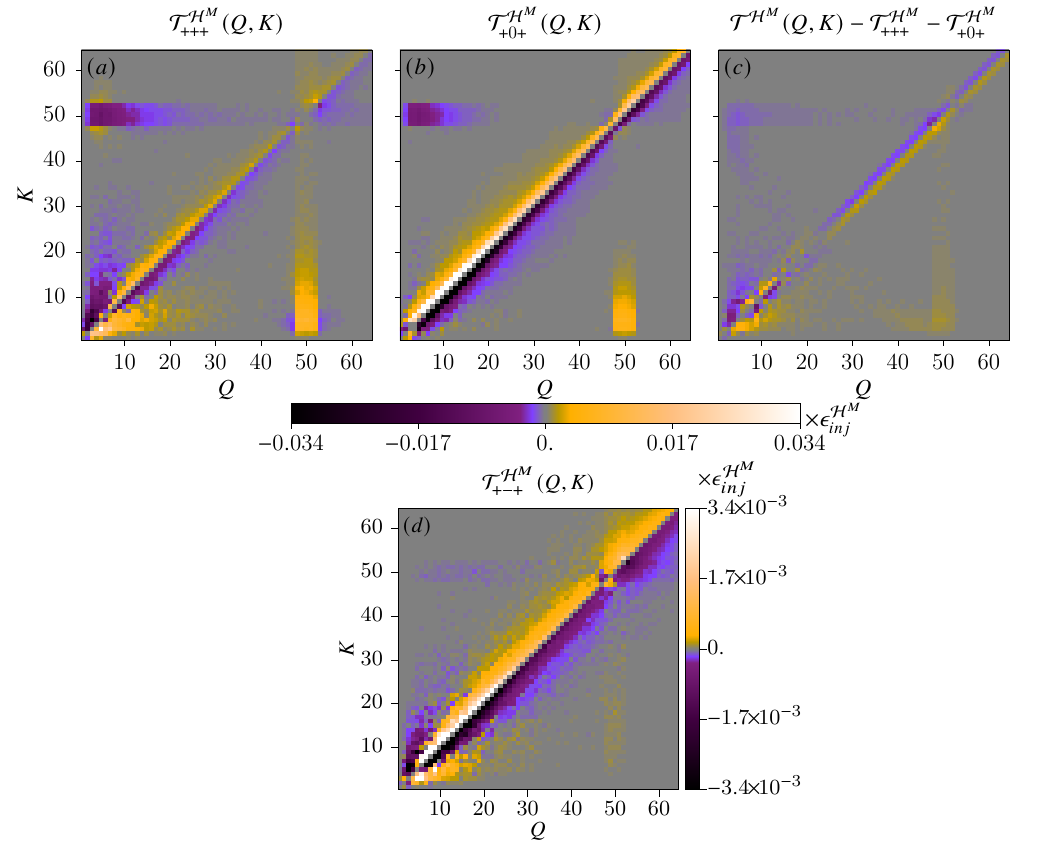}{Dominant helical transfer terms (\eq \eqref{eq:tsfmhlhel}) from \fig \ref{fig:tsfhmqk}: $(a)$ mediated by the like-signed helical velocity field and $(b)$ by the compressive part of the velocity field. The remaining transfer terms $(c)$ are relatively small, e.g. the one mediated by the oppositely-signed helical velocity field $(d)$. Note the order of magnitude difference between the color bars.}{fig:tsfhmqkhel}

Which helical component plays a role for which of the three LDT, LIT and NLIT features can be explained by considering the geometric helical transfer moduli (\eqs \eqref{eq:Gincomp}-\eqref{eq:Gcomp}, plotted in \fig \ref{fig:Gplot}). Comparing \figs \ref{fig:tsfhmqkhel} and \ref{fig:Gplot}, one observes that:

\begin{enumerate}
\item The LIT is mediated by the like-signed helical velocity field (and, to a smaller extent, by the oppositely-signed helical one), but the compressive part of the velocity field plays no role in it. Indeed, for these local transfers $k\approx q \approx p$ mediated by the intermediate scale velocity field, $G^{+++}$ presents a maximum, whereas $G^{+-+}$ is significantly smaller and $G^{+0+}$ vanishes.
\item The NLIT is mediated by both the like-signed helical velocity field and the compressive velocity field, consistently with high values of $G^{+++}$ and $G^{+0+}$ for small velocity scales: $q \ll p \approx k$ and $k \ll p \approx q$. The contribution from the compressive velocity field is more nonlocal (compare \fig \ref{fig:tsfhmqkhel}.$(a)$ with \ref{fig:tsfhmqkhel}.$(b)$) since $G^{+0+}$ vanishes for $k\approx q$. The oppositely-signed helical velocity field plays a negligible role since $G^{+-+}$ vanishes for $p \gg k$ and $p \gg q$. 
\item The LDT, which corresponds to the large-scale velocity field $p\ll k \approx q$, is mediated by all helical components. The compressive velocity field is however geometrically favored, since $(i)$ on the $q=k\pm p$ lines, $G^{+0+}=2$ while $G^{+++}=G^{+-+}=0$ and $(ii)$ the $q=k$ line, where $G^{+++}$ and $G^{+-+}$ are high at low $p$, does not play a role in the magnetic helicity transfer since $\TsfMhl(K,K)=0$. Both solenoidal contributions are of the same order of magnitude for the LDT since $G^{+++}\approx G^{+-+}$ for $p\ll q\approx k$.
\end{enumerate}

\newfig{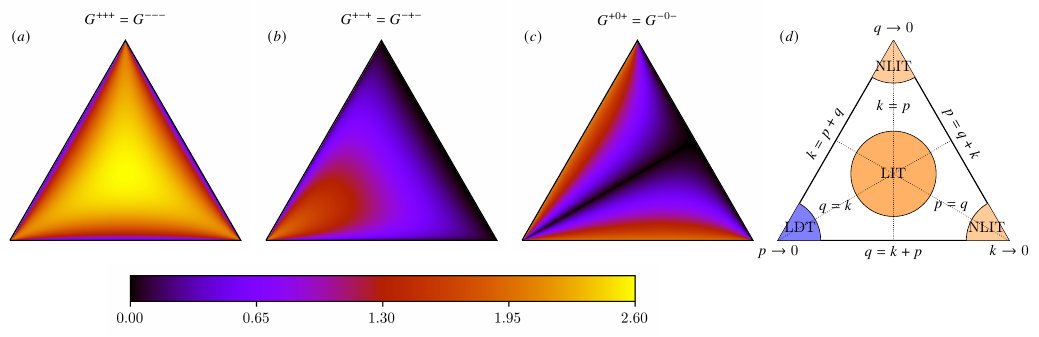}{$(a),(b),(c):$ Moduli of the geometric helical factor (\eqs \eqref{eq:Gincomp}-\eqref{eq:Gcomp}), for the homochiral terms $\hsK=\hsQ$. $(d)$ Key to read the figures: the triangles are delimited by the lines $k=p+q$, $p=q+k$ and $q=k+p$, where the transfer is from magnetic mode $\vq$ to magnetic mode $\vk$ mediated by the velocity mode $\vp$. The region $p \ll q\approx k$ corresponds to the large-scale velocity field, i.e. the LDT, the region $p\approx k \approx q$ to the intermediate scale local transfers, i.e. the LIT and the regions $p\approx k \gg q$ and $p\approx q \gg k$ to the small-scale velocity field, i.e. the NLIT.}{fig:Gplot}

The above results confirm and extend those from incompressible MHD. Through a comparison of $G^{+++}$ and $G^{+-+}$, it has been predicted that the transfers mediated by the like-signed helical velocity field should be stronger and more nonlocal than those mediated by the oppositely-signed helical one \citep{LSM17}.

A more detailed analysis of the other simulations with higher and lower Mach number \citep{TEM21b}: M01s, M11s and M1c confirms the results reviewed here.

\section{Summary and conclusion}
\label{sec:conclusion}

\newfig{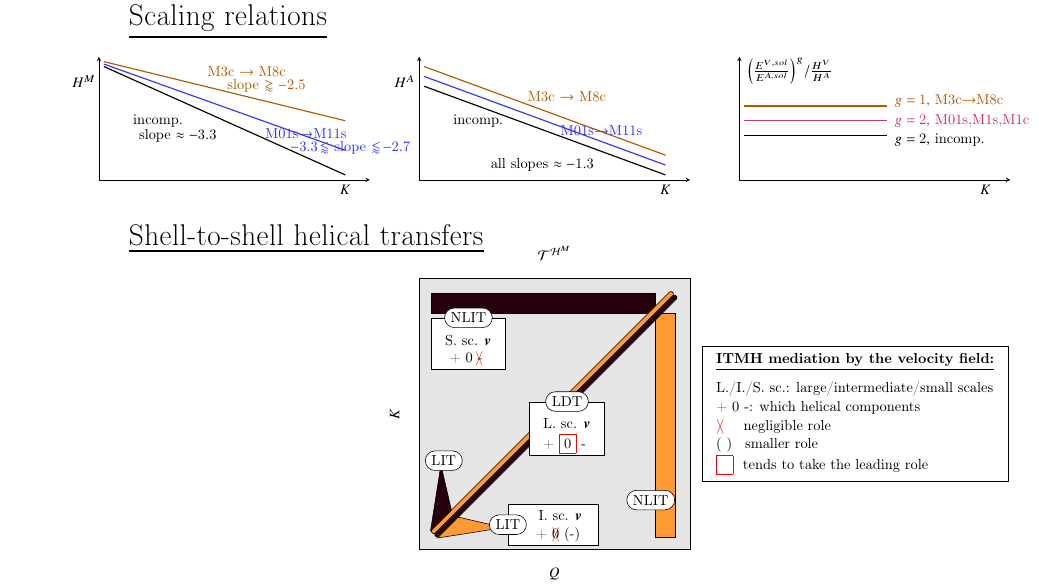}{Graphical visualisation of the main results.}{fig:conclusion}

The main results are summarized visually in \fig \ref{fig:conclusion}: 
\begin{enumerate}
\item The magnetic helicity spectra tend to become flatter with higher compressibility. The departure from the incompressible scaling depends more on the driving type than on the turbulent Mach number. This suggests that for astrophysical systems of interest, compressible effects are important at already relatively low Mach numbers, provided the turbulence drivers are compressive enough.
\item The Alfv\'enic helicity (the helicity of the Alfv\'en velocity $\Fmag/\sqrt{\rho}$) presents a scaling law close to the incompressible one, independently on the driving mechanism, in a wide range of Mach numbers.
\item The dynamic balance between shear and twisting (Alfv\'enic balance) can be extended to supersonic MHD when considering the solenoidal part of the specific kinetic energy spectra and the Alfv\'en velocity in place of the magnetic field.
\item Three phenomena occur in the global picture of the \MHIT:
	\begin{itemize}
	\item a local direct transfer (LDT) mediated by the large-scale velocity field. Its compressive part is geometrically favored. The like-signed and oppositely-signed helical part take similar roles.
	\item a local inverse transfer (LIT) mediated by the intermediate-scale velocity field. The like-signed helical part plays the main role, the oppositely-signed a subdominant role. The importance of the compressive velocity field for this transfer is negligible.
	\item a nonlocal inverse transfer (NLIT) mediated by the small-scale velocity field. The like-signed helical part as well as the compressive part play the dominant roles, whereas the oppositely-signed helical part plays a negligible role.
	\end{itemize}
\end{enumerate}

These results have been gained from direct numerical simulations of the ideal supersonic isothermal magnetohydrodynamics equations, subject to both a large-scale mechanical forcing (either purely solenoidal or purely compressive) and a small-scale helical electromotive driving \citep{TEM21a,TEM21b}. The use of a fourth-order numerical scheme allows good accuracy even at the moderate $512^3$ resolution. The turbulent Mach numbers span from subsonic to about 10. Even though they are obtained apart from a statistically stationary state, so that transient effects cannot be excluded, the confidence in the main results is relatively high because $(i)$ they have been confirmed at higher resolution ($1024^3$) and $(ii)$ their domain of validity spans over a wide range of turbulent Mach numbers.

In the presented setting, magnetic helicity of one sign dominates the system at all scales. However, in geophysical and astrophysical systems of interest, such as dynamos, both signs are expected to be present. This could lead to other scaling laws and to a greater importance of the heterochiral transfer terms, which are more difficult to interpret (cf \cite{TEI20}, sections 7.1 and C.2). Nevertheless, these first results in this simpler setting should be helpful to interpret future investigations.

\bibliographystyle{abbrv}%
\bibliography{biblio}%
\end{document}

%% file: includes/newcommands.tex
\input{nc_num}
\input{nc_fourier}

\input{nc_tilde}
\input{nc_arrows}
\input{nc_domain}
\input{nc_eq}

\input{nc_vec}

\input{nc_bar}
\input{nc_misc}

\input{nc_physics}

\input{nc_experiments}
\input{nc_tikz}

%% file: includes/nc_fourier.tex
\newcommand{\vhf}{\vec{\hat {f}}}
\newcommand{\vhg}{\vec{\hat {g}}}

\newcommand{\vhhk}{\vec{{\hat{h}}^0_{\vk}}}
\newcommand{\vhhp}{\vec{{\hat{h}}^+_{\vk}}}
\newcommand{\vhhm}{\vec{{\hat{h}}^-_{\vk}}}
\newcommand{\vhhpm}{\vec{{\hat{h}}^\pm_{\vk}}}
\newcommand{\vhhs}[2]{\vec{{{\hat{h}}}^{#1}_{#2}}}

\newcommand{\vhe}{{\vec{\hat{e}}}}

%% file: includes/nc_eq.tex
\newcommand{\beqa}{\begin{eqnarray}}
\newcommand{\eeqa}{\end{eqnarray}}
\newcommand{\beq}{\begin{equation}}
\newcommand{\eeq}{\end{equation}}

\newcommand{\beqal}{\begin{equation}\begin{aligned}}
\newcommand{\eeqal}{\end{aligned}\end{equation}}

\newcommand{\pat}{\partial_t}

%% file: includes/nc_vec.tex
\newcommand{\vf}{\vec{{f}}}

\newcommand{\vg}{\vec{{g}}}

\newcommand{\vk}{{\vec{k}}}

\newcommand{\vN}{\vec{N}}

\newcommand{\vom}{\vec {\omega}}

\newcommand{\vp}{\vec{p}}
\newcommand{\vq}{\vec{q}}

\newcommand{\vW}{\vec{W}}

%% file: includes/nc_misc.tex
\newcommand{\bit}{\begin{itemize}}
\newcommand{\eit}{\end{itemize}}

%% file: includes/nc_physics.tex

\newcommand{\helP}{+}
\newcommand{\helC}{0}
\newcommand{\helN}{-}

\newcommand{\neutrfield}{\vf}
\newcommand{\neutrfieldB}{\vg}
\newcommand{\neutrfieldF}{\vhf}
\newcommand{\neutrfieldBF}{\vhg}

\newcommand{\solexpo}{sol}

\newcommand{\rmsM}{{\cal{M}}} 
\newcommand{\kspec}{K} 



\newcommand{\hel}{{\cal{H}}}
\newcommand{\mhel}{{\hel^M}} 

\newcommand{\ahelF}{H^A} 
\newcommand{\ahelFK}{H^A_K} 
\newcommand{\jhelF}{H^J} 
\newcommand{\jhelFK}{H^J_K} 
\newcommand{\mhelF}{H^M} 
\newcommand{\mhelFK}{\mhelF_\kspec} 
\newcommand{\khelF}{H^V} 

\newcommand{\ealfFsol}{E^{A,\solexpo}} 
\newcommand{\emagF}{E^M} 





\newcommand{\sekinF}{E^V} 

\newcommand{\sekinFsol}{E^{V,\solexpo}} 


\newcommand{\famp}{F_0}

\newcommand{\fOUF}{\vec{F^{OU}}}
\newcommand{\fOUFk}{\vec{ F^{OU}_{\vk}}}
\newcommand{\fat}{t_{auto}} 
\newcommand{\Lbox}{L} 

\newcommand{\Einj}{{\epsilon^{K}_{inj}}} 
\newcommand{\EMinj}{{\epsilon^{M}_{inj}}} 
\newcommand{\mhelinj}{{\epsilon^{\mhel}_{inj}}} 

 %
 %



\newcommand{\FmagL}{b} 
\newcommand{\Fmag}{{\vec{\FmagL}}} 
\newcommand{\FmagK}{\Fmag_K} 

\newcommand{\FmagQ}{\Fmag_Q} 
\newcommand{\FmagLF}{{{}\hat{b}}} 
\newcommand{\FmagF}{\vec{\FmagLF}} 
\newcommand{\FmagFL}{\FmagL} 
\newcommand{\FmagAL}{a} 
\newcommand{\FmagA}{{\vec{\FmagAL}}} 
\newcommand{\FmagAK}{\FmagA_K} 

\newcommand{\FvelL}{v} 
\newcommand{\Fvel}{{\vec{\FvelL}}} 
\newcommand{\FvelP}{\Fvel_P} 

\newcommand{\FvelLF}{{{}\hat{v}}} 
\newcommand{\FvelF}{\vec{\FvelLF}} 
\newcommand{\FvelFL}{\FvelL} 


\newcommand{\FmagXX}[2]{\Fmag^{#1}_{#2}}
\newcommand{\FvelXX}[2]{\Fvel^{#1}_{#2}}

\newcommand{\helsign}{s}

\newcommand{\helsignk}{\helsign_k}
\newcommand{\helsignp}{\helsign_p}
\newcommand{\helsignq}{\helsign_q}

\newcommand{\hsk}{\helsignk}
\newcommand{\hsp}{\helsignp}
\newcommand{\hsq}{\helsignq}

\newcommand{\helsignK}{\helsign_K}
\newcommand{\helsignP}{\helsign_P}
\newcommand{\helsignQ}{\helsign_Q}

\newcommand{\hsK}{\helsignK}
\newcommand{\hsP}{\helsignP}
\newcommand{\hsQ}{\helsignQ}

\newcommand{\triadgeom}{g}

\newcommand{\triadgeomkpq}{\triadgeom^{\helsignk,\helsignp,\helsignq}_{k,p,q}}

\newcommand{\FvelA}{{\vec{\FvelL_A}}} 


\newcommand{\cs}{c_s} 

\newcommand{\specw}{\zeta} 
\newcommand{\projk}{{\cal{P}^{\specw}}} 
\newcommand{\projkij}{{{\cal{P}}^{\specw}_{ij}}} 
\newcommand{\projkT}{\underline{\projk}} 


\newcommand{\kone}{\kappa} 



\newcommand{\IscHm}{{\cal{I}}_{\mhel}}

\newcommand{\mhelfracFor}{h_f}

\newcommand{\rhoz}{\rho_0}

\newcommand{\mhlkinj}{\kspec^M_f}

\newcommand{\PPP}{\helP\helP\helP}
\newcommand{\PCP}{\helP\helC\helP}
\newcommand{\PNP}{\helP\helN\helP}


\newcommand{\Tsf}{{\cal{T}}}

\newcommand{\TsfMhl}{\Tsf^{\mhel}}

\newcommand{\TsfMhlQPK}{\TsfMhl(Q,P,K)}

\newcommand{\TsfMhlHEL}[1]{\Tsf^{\mhel}_{#1}}

\newcommand{\TsfMhlXXX}[1]{\TsfMhlHEL{#1}}










\newcommand{\powerspec}{{\cal{P}}}

\newcommand{\powerspecsol}{\powerspec^{\solexpo}}
\newcommand{\productspec}{{\cal{Q}}}



\newcommand{\ForceV}{\vec{f}_V}

\newcommand{\ForceM}{\vec{f}_M}

\newcommand{\mathI}{\vec I}
\newcommand{\upi}{\uppi}
\renewcommand{\vec}[1]{\boldsymbol{#1}}

\newcommand{\volavg}[1]{\langle #1 \rangle}

%% file: aguchapter.bbl
\begin{thebibliography}{10}

\bibitem{ALE17}
A.~Alexakis.
\newblock Helically decomposed turbulence.
\newblock {\em Journal of Fluid Mechanics 812}, pages 752--770, 2017.

\bibitem{ALB18}
A.~Alexakis and L.~Biferale.
\newblock Cascades and transitions in turbulent flows.
\newblock {\em Physics Reports 767-769}, pages 1--101, 2018.

\bibitem{AMP06}
A.~Alexakis, P.~D. Mininni, and A.~Pouquet.
\newblock On the inverse cascade of magnetic helicity.
\newblock {\em The Astrophysical Journal 640}, pages 335--343, 2006.

\bibitem{ALU13}
H.~Aluie.
\newblock Scale decomposition in compressible turbulence.
\newblock {\em Physica D 247}, pages 54--65, 2013.

\bibitem{ALU17}
H.~Aluie.
\newblock Coarse-grained incompressible magnetohydrodynamics: analyzing the
  turbulent cascades.
\newblock {\em New Journal of Physics 19, 025008}, 2017.

\bibitem{BAP99}
D.~Balsara and A.~Pouquet.
\newblock The formation of large-scale structures in supersonic
  magnetohydrodynamic flows.
\newblock {\em Physics of Plasmas Vol. 6 No. 1}, pages 89--99, 1999.

\bibitem{BAL12}
D.~S. Balsara.
\newblock Self-adjusting, positivity preserving high order schemes for
  hydrodynamics and magnetohydrodynamics.
\newblock {\em Journal of Computational Physics 231}, pages 7504--7517, 2012.

\bibitem{BGS16}
D.~S. Balsara, S.~Garain, and C.-W. Shu.
\newblock An efficient class of {WENO} schemes with adaptive order.
\newblock {\em Journal of Computational Physics 326}, pages 780--804, 2016.

\bibitem{BER97}
M.~A. Berger.
\newblock Magnetic helicity in a periodic domain.
\newblock {\em Journal of Geophysical Research 102 No. A2}, pages 2637--2644,
  1997.

\bibitem{BER99}
M.~A. Berger.
\newblock Introduction to magnetic helicity.
\newblock {\em Plasma Physics and Controlled Fusion 41}, pages B167--B175,
  1999.

\bibitem{BEF84}
M.~A. Berger and G.~B. Field.
\newblock The topological properties of magnetic helicity.
\newblock {\em Journal of Fluid Mechanics 147}, pages 133--148, 1984.

\bibitem{BMT12}
L.~Biferale, S.~Musacchio, and F.~Toschi.
\newblock Inverse energy cascade in three-dimensional isotropic turbulence.
\newblock {\em Physical Review Letters 108, 164501}, 2012.

\bibitem{BIT13}
L.~Biferale and E.~S. Titi.
\newblock On the global regularity of a helical-decimated version of the {3D}
  {N}avier-{S}tokes equations.
\newblock {\em Journal of Statistical Physics 151}, pages 1089--1098, 2013.

\bibitem{BLA15}
E.~G. Blackman.
\newblock Magnetic helicity and large scale magnetic fields: A primer.
\newblock {\em Space Science Reviews 188}, pages 59--91, 2015.

\bibitem{BRA01}
A.~Brandenburg.
\newblock The inverse cascade and nonlinear alpha-effect in simulations of
  isotropic helical hydromagnetic turbulence.
\newblock {\em The Astrophysical Journal 550}, pages 824--840, 2001.

\bibitem{BRS05}
A.~Brandenburg and K.~Subramanian.
\newblock Astrophysical magnetic fields and nonlinear dynamo theory.
\newblock {\em Physics Reports 417}, pages 1--209, 2005.

\bibitem{BUH14}
P.~Buchm\"{u}ller and C.~Helzel.
\newblock Improved accuracy of high-order {WENO} finite volume methods on
  cartesian grids.
\newblock {\em Journal of Scientific Computing 61}, pages 343--368, 2014.

\bibitem{CHQ88}
C.~Canuto, M.~Y. Hussaini, A.~Quarteroni, and T.~A. Zang.
\newblock {\em Spectral Methods in Fluid Dynamics}.
\newblock Springer-Verlag, 1988.

\bibitem{CAO07}
B.~W. Caroll and D.~A. Ostlie.
\newblock {\em An Introduction to Modern Astrophysics, second edition}.
\newblock Pearson Education, 2007.

\bibitem{CHB01}
M.~Christensson, M.~Hindmarsh, and A.~Brandenburg.
\newblock Inverse cascade in decaying three-dimensional magnetohydrodynamic
  turbulence.
\newblock {\em Physical Review E 64, 056405}, 2001.

\bibitem{COM88}
P.~Constantin and A.~Majda.
\newblock The {B}eltrami spectrum for incompressible fluid flows.
\newblock {\em Communications in Mathematical Physics 115}, pages 435--456,
  1988.

\bibitem{ELS04}
B.~G. Elmegreen and J.~Scalo.
\newblock Interstellar turbulence {I}: Observations.
\newblock {\em Annual Review of Astronomy and Astrophysics 42}, pages 211--273,
  2004.

\bibitem{ELS56}
W.~M. Els\"{a}sser.
\newblock Hydromagnetic dynamo theory.
\newblock {\em Reviews of Modern Physics 28, No. 2}, pages 135--163, 1956.

\bibitem{FED13}
C.~Federrath.
\newblock On the universality of supersonic turbulence.
\newblock {\em Monthly Notices of the Royal Astronomical Society 436}, pages
  1245--1257, 2013.

\bibitem{FPL75}
U.~Frisch, A.~Pouquet, J.~L\'{e}orat, and A.~Mazure.
\newblock Possibility of an inverse cascade of magnetic helicity in
  magnetohydrodynamic turbulence.
\newblock {\em Journal of Fluid Mechanics 68 Part 4}, pages 769--778, 1975.

\bibitem{GAB11}
S.~Galtier and S.~Banerjee.
\newblock Exact relation for correlation functions in compressible isothermal
  turbulence.
\newblock {\em Physical Review Letters 107, 134501}, 2011.

\bibitem{GMP11}
J.~Graham, P.~D. Mininni, and A.~Pouquet.
\newblock High {R}eynolds number magnetohydrodynamic turbulence using a
  {L}agrangian model.
\newblock {\em Physical Review E 84, 016314}, 2011.

\bibitem{HEO87}
A.~Harten, B.~Engquist, S.~Osher, and S.~R. Chakravarthy.
\newblock Uniformly high order accurate essentially non-oscillatory schemes,
  {III}.
\newblock {\em Journal of Computational Physics 71}, pages 231--303, 1987.

\bibitem{JIS96}
G.-S. Jiang and C.-W. Shu.
\newblock Efficient implementation of weighted {ENO} schemes.
\newblock {\em Journal of Computational Physics 126}, pages 202--228, 1996.

\bibitem{KET08}
D.~I. Ketcheson.
\newblock Highly efficient strong stability-preserving {Runge-Kutta} methods
  with low-storage implementations.
\newblock {\em Society for Industrial and Applied Mathematics Journal on
  Scientific Computing 30, No. 4}, pages 2113--2136, 2008.

\bibitem{KIV95}
M.~G. Kivelson.
\newblock Physics of space plasmas.
\newblock In M.~G. Kivelson and C.~T. Russell, editors, {\em Introduction to
  Space Physics}, chapter~2. Cambridge University Press, 1995.

\bibitem{KWN13}
A.~G. Kritsuk, R.~Wagner, and M.~L. Norman.
\newblock Energy cascade and scaling in supersonic isothermal turbulence.
\newblock {\em Journal of Fluid Mechanics Vol. 729, R1}, 2013.

\bibitem{KUR96}
A.~Kumar and D.~M. Rust.
\newblock Interplanetary magnetic clouds, helicity conservation, and
  current-core flux-ropes.
\newblock {\em Journal of Geophysical Research 101}, pages 667--684, 1996.

\bibitem{LPC09}
T.~Lessinnes, F.~Plunian, and D.~Carati.
\newblock Helical shell models for {MHD}.
\newblock {\em Theoretical and Computational Fluid Dynamics 23}, pages
  439--450, 2009.

\bibitem{LBM16}
M.~Linkmann, A.~Berera, M.~McKay, and J.~J\"{a}ger.
\newblock Helical mode interactions and spectral transfer processes in
  magnetohydrodynamic turbulence.
\newblock {\em Journal of Fluid Mechanics 791}, pages 61--96, 2016.

\bibitem{LID16}
M.~Linkmann and V.~Dallas.
\newblock Large-scale dynamics of magnetic helicity.
\newblock {\em Physical Review E 94, 053209}, 2016.

\bibitem{LID17}
M.~Linkmann and V.~Dallas.
\newblock Triad interactions and the bidirectional turbulent cascade of
  magnetic helicity.
\newblock {\em Physical Review Fluids 2, 054605}, 2017.

\bibitem{LSM17}
M.~Linkmann, G.~Sahoo, M.~McKay, A.~Berera, and L.~Biferale.
\newblock Effects of magnetic and kinetic helicities on the growth of magnetic
  fields in laminar and turbulent flows by helical {F}ourier decomposition.
\newblock {\em The Astrophysical Journal 836:26}, 2017.

\bibitem{LOC94}
X.-D. Liu, S.~Osher, and T.~Chan.
\newblock Weighted essentially non-oscillatory schemes.
\newblock {\em Journal of Computational Physics 115}, pages 200--212, 1994.

\bibitem{LOW94}
B.~C. Low.
\newblock Magnetohydrodynamic processes in the solar corona: Flares, coronal
  mass ejections, and magnetic helicity.
\newblock {\em Physics of Plasmas 1}, pages 1684--1690, 1994.

\bibitem{COC11}
P.~McCorquodale and P.~Colella.
\newblock A high-order finite-volume method for conservation laws on locally
  refined grids.
\newblock {\em Communications in Applied Mathematics and Computational Science
  6, No. 1}, pages 1--25, 2011.

\bibitem{MFP81}
M.~Meneguzzi, U.~Frisch, and A.~Pouquet.
\newblock Helical and nonhelical turbulent dynamos.
\newblock {\em Physical Review Letters 47, No. 15}, pages 1060--1064, 1981.

\bibitem{MIP09}
P.~D. Mininni and A.~Pouquet.
\newblock Finite dissipation and intermittency in magnetohydrodynamics.
\newblock {\em Physical Review E 80, 025401}, 2009.

\bibitem{MOF69}
H.~K. Moffatt.
\newblock The degree of knottedness of tangled vortex lines.
\newblock {\em Journal of Fluid Mechanics 35}, pages 117--129, 1969.

\bibitem{MUM10}
W.-C. M\"{u}ller and S.~K. Malapaka.
\newblock Understanding nonlinear cascades in magnetohydrodynamic turbulence by
  statistical closure theory.
\newblock {\em ASP Conference Series 429}, 2010.
\newblock Note: $H^K$ and $H^M$ have been erroneously exchanged starting eq.
  (4).

\bibitem{MMB12}
W.-C. M\"{u}ller, S.~K. Malapaka, and A.~Busse.
\newblock Inverse cascade of magnetic helicity in magnetohydrodynamic
  turbulence.
\newblock {\em Physical Review E 85, 015302}, 2012.

\bibitem{PSV19}
F.~Plunian, R.~Stepanov, and M.~K. Verma.
\newblock On uniqueness of transfer rates in magnetohydrodynamic turbulence.
\newblock {\em Journal of Plasma Physics 85}, 2019.

\bibitem{PFL76}
A.~Pouquet, U.~Frisch, and J.~L\'{e}orat.
\newblock Strong {MHD} helical turbulence and the nonlinear dynamo effect.
\newblock {\em Journal of Fluid Mechanics 77, part 2}, pages 321--354, 1976.

\bibitem{POP78}
A.~Pouquet and G.~S. Patterson.
\newblock Numerical simulation of helical magnetohydrodynamic turbulence.
\newblock {\em Journal of Fluid Mechanics 85, part 2}, pages 305--323, 1978.

\bibitem{SHU09}
C.-W. Shu.
\newblock High order weighted essentially nonoscillatory schemes for convection
  dominated problems.
\newblock {\em Society for Industrial and Applied Mathematics Review Vol. 51,
  No. 1}, pages 82--126, 2009.

\bibitem{SFM15}
R.~Stepanov, P.~Frick, and I.~Mizeva.
\newblock Joint inverse cascade of magnetic energy and magnetic energy in {MHD}
  turbulence.
\newblock {\em The Astrophysical Journal Letters 798}, 2015.

\bibitem{TEI20}
J.-M. Teissier.
\newblock {\em Magnetic helicity inverse transfer in isothermal supersonic
  magnetohydrodynamic turbulence}.
\newblock PhD thesis, Technische Universit\"{a}t Berlin, 2020.

\bibitem{TEM21b}
J.-M. Teissier and W.-C. M\"{u}ller.
\newblock Inverse transfer of magnetic helicity in direct numerical simulations
  of compressible isothermal turbulence: helical transfers.
\newblock {\em Journal of Fluid Mechanics 921, A7}, 2021.
\newblock Note: the normalization factors of 1/(2$\sqrt{2}$) or 1/2 in eqs.
  5.5, 5.6, A5, A6 and fig. 8 are missing.

\bibitem{TEM21a}
J.-M. Teissier and W.-C. M\"{u}ller.
\newblock Inverse transfer of magnetic helicity in direct numerical simulations
  of compressible isothermal turbulence: scaling laws.
\newblock {\em Journal of Fluid Mechanics 915, A23}, 2021.

\bibitem{VTH19}
P.~S. Verma, J.-M. Teissier, O.~Henze, and W.-C. M\"{u}ller.
\newblock Fourth-order accurate finite-volume {CWENO} scheme for astrophysical
  {MHD} problems.
\newblock {\em Monthly Notices of the Royal Astronomical Society 482}, pages
  416--437, 2019.

\bibitem{VIC01}
E.~T. Vishniac and J.~Cho.
\newblock Magnetic helicity conservation and astrophysical dynamos.
\newblock {\em The Astrophysical Journal 550}, pages 752--760, 2001.

\bibitem{WAL92}
F.~Waleffe.
\newblock The nature of triad interactions in homogeneous turbulence.
\newblock {\em Physics of Fluids A: Fluid Dynamics 4}, pages 350--363, 1992.

\bibitem{WOL58a}
L.~Woltjer.
\newblock A theorem on force-free magnetic fields.
\newblock {\em Proceedings of the National Academy of Sciences 44, No. 6},
  pages 489--491, 1958.

\bibitem{WSH19}
K.~Wu and C.-W. Shu.
\newblock Provably positive high-order schemes for ideal magnetohydrodynamics:
  analysis on general meshes.
\newblock {\em Numerische Mathematik 142}, pages 995--1047, 2019.

\end{thebibliography}
